# Phase-stabilized, 1.5-W frequency comb at 2.8 to 4.8 μm

Florian Adler,[1,*] Kevin C. Cossel,[1] Michael J. Thorpe,[1] Ingmar Hartl,[2] Martin E. Fermann,[2] and Jun Ye[1]

[1]*JILA, National Institute of Standards and Technology and University of Colorado Department of Physics, University of Colorado, Boulder, Colorado 80309-0440, USA*

[2]*IMRA America Inc., 1044 Woodridge Avenue, Ann Arbor, Michigan 48105, USA*

[*]*Corresponding author: fadler@jila.colorado.edu*

We present a high-power optical parametric oscillator-based frequency comb in the mid-infrared wavelength region using periodically poled lithium niobate. The system is synchronously pumped by a 10-W femtosecond Yb:fiber laser centered at 1.07 μm and is singly resonant for the signal. The idler (signal) wavelength can be continuously tuned from 2.8 to 4.8 μm (1.76 to 1.37 μm) with a simultaneous bandwidth as high as 0.3 μm and a maximum average idler output power of 1.50 W. We also demonstrate the performance of the stabilized comb by recording the heterodyne beat with a narrow-linewidth diode laser. This OPO is an ideal source for frequency comb spectroscopy in the mid-IR.  2009 Optical Society of America

*OCIS codes:* 190.4970, 320.7160, 140.7090.

Over the past years, frequency combs have become attractive laser sources for a variety of applications – the latest of which is the direct use of comb sources for precision spectroscopy. Frequency combs provide a unique combination of wide wavelength coverage and high spectral resolution, and therefore they allow for simultaneous, precise and rapid scanning of wide regions of interest [1–6]. Furthermore, the spectroscopic sensitivity may be massively increased by efficiently coupling the comb to a high-finesse enhancement cavity. Consequently, cavity-enhanced frequency comb spectroscopy (CE-DFCS) [7] proved very successful for detection of trace gases such as bio-markers in human breath [8]. Recently, a first application in cold molecule spectroscopy has also been demonstrated [9].

Despite its success, CE-DFCS has so far not been able to exploit its full potential. Practical comb sources are mostly limited to Ti:sapphire and Er:fiber lasers and therefore to the near-infrared spectral region of less than 2 µm; however, the fundamental ro-vibrational absorption bands of most molecules lie in the mid-infrared between 3 and 12 µm. Therefore, presently used frequency combs have to rely on weaker overtone vibrations, which compromises the detection sensitivity by about two to three orders of magnitude. There are two common strategies to generate a mid-infrared frequency comb: difference-frequency generation (DFG) and optical parametric oscillators (OPO). Employing the first method, it is easier to achieve a stable IR comb, especially when the DFG light is generated from the same source. The power level that can be provided, however, typically does not exceed 1 mW [10–13]. In contrast, femtosecond mid-IR OPOs based on periodically poled lithium niobate (PPLN) [14,15] are able to provide higher output powers on the order of several hundred milliwatts; a stable idler frequency comb, however, has only been demonstrated for wavelengths below 2.5 µm [16,17].

In this Letter we present a fiber-laser pumped OPO based on a fan-out MgO-doped PPLN crystal that provides a high-power frequency comb in the mid-infrared spectral region. We generate up to 300-nm-wide idler spectra, which are easily and continuously tunable from 2.8 to 4.8 µm with an average power of up to 1.5 W. The corresponding signal covers the spectral range of 1.37–1.76 µm. We also present a scheme for full electronic stabilization of the mid-IR frequency comb and demonstrate future capabilities by recording the optical beat note of the doubled idler signal with a narrow-linewidth continuous-wave diode laser at 1.545 µm.

A schematic of our setup is depicted in Fig. 1(a). The OPO is designed with a 5-mirror linear cavity, which is singly resonant for the signal. Its length is matched to the 136 MHz repetition rate of the pump laser (corresponding to a 110-cm OPO cavity length). The amplified Yb:fiber pump laser delivers 100-fs-pulses with a maximum average power of 10 W at a center wavelength of 1.07 µm [18] and is coupled into the cavity through mirror M2 (R < 3%). The OPO crystal is a 7-mm-long MgO-doped PPLN with fan-out grating structure, *i.e.*, the poling period varies linearly over the width of the crystal [19]. Therefore, the OPO's wavelength is tuned fast and easily by simply translating the PPLN. The shortest (25.5 µm) and longest (32.5 µm) poling periods are designed to provide first order quasi-phase-matching for idler wavelengths of 5 µm and 3 µm, respectively. As a precaution to prevent damage from the intense pump beam, the crystal is heated to a temperature of 160 ºF. Therefore, the idler output is expected to be shifted to somewhat shorter wavelengths compared to the original design. As we currently do not require the signal output, all cavity mirrors (M1–M5) in the present setup are highly reflective for 1.35–1.80 µm. Mirror M4 couples out both pump and idler with an average transmission of 95%. Due to the extremely high nonlinearity of the MgO:PPLN crystal, a variety of generally non-phase-matched mixing signals is emitted along with signal ($s$) and idler ($i$). Fig. 1(b) shows a spectrum of the light leaking through end mirror M1 recorded with an optical spectrum analyzer (OSA), along with the corresponding idler spectrum. The most prominent mixing signals are the sum-frequencies of pump and signal ($p+s$), pump and idler ($p+i$), frequency-doubled pump ($2p$) and signal ($2s$). Although these wavelengths are only by-products from the OPO process, they exhibit powers of several milliwatts. A dichroic mirror (DM) after M4 separates the idler from the residual light, which is used to stabilize the mid-IR comb.

The powerful pump laser and the high nonlinearity of the PPLN lead to a high conversion efficiency into the infrared over a wide tuning range. Fig. 2 shows the idler output power (black triangles, left axis) and photon conversion efficiency (grey circles, right axis) measured behind M4 for different center wavelengths. The pump laser is operated at 8.47 W at all crystal positions. The maximum idler power (1.42 W) is obtained at 3.19 µm, whereas the highest photon conversion efficiency of 51% is observed at an idler wavelength of 3.59 µm, which results in 1.29 W of average power. Most notably, a power of at least 1 W is maintained over a wide tuning range of more than 1 µm. The drops at short and long idler wavelengths agree well with the expected trend due to the mirror reflectivities (grey dashed line). In order to extract the OPO threshold and slope efficiency, we plot the mid-IR output power against pump power at one specific crystal position ($\lambda_{idler}$ = 3.01 µm), as shown in the inset of Fig. 2. The threshold pump power is determined as 1.7 W, which is somewhat higher than expected from a femtosecond OPO. One has to consider, however, that OPO cavity and pump laser are not dispersion managed because short pulse duration is neither required nor desirable for the targeted linear spectroscopy applications, and that the strong pump light is not tightly focused into the crystal to prevent material damage. The slope efficiency starts off with 0.29 and reduces to 0.16 at about 4 W of pump power due to the onset of a competing conversion process. Most notably, we do not observe any indication of nonlinear losses with up to 9.17 W of pump power, where we measure 1.50 W of average idler power. In addition, spectral data is recorded with a monochromator and a liquid-nitrogen-cooled InSb detector, as depicted in Fig. 3. The center wavelength is tuned solely by translating the crystal and ranges from 2.8 µm to 4.8 µm. The widest simultaneous bandwidth (full width at half maximum) is 0.3 µm, obtained at 3.26 µm.

Finally, we demonstrate the OPO's capability to serve as mid-IR frequency comb by stabilizing the idler output. Its *n*-th comb tooth may be described by the usual frequency comb equation $v_n^i = f_0^i + n \cdot f_{rep}^i$, where $f_0^x$ and $f_{rep}^x$ stand for the comb offset and repetition frequency, respectively: *x* can be *i* (idler), *s* (signal) or *p* (pump), and *n* is an integer number on the order of $10^5$–$10^6$. The comb offset frequencies of the OPO, $f_0^p$, $f_0^s$ and $f_0^i$ are connected by the relation $f_0^p = f_0^s + f_0^i$, which means that the temporal phase shift between *s* and *i* is the same after each round trip when the cavity length is exactly matched to $f_{rep}^p$ (and $f_0^p$ is stable).

The stabilization scheme is implemented by exploiting the parasitic mixing signals generated by the PPLN crystal (see Fig. 1). In general we utilize $p+s$ and $p+i$, which are separated by a dichroic mirror (DM) and focused onto Si detectors (PD1 and PD2, respectively). To generate a useful beat note, we pick off a small portion of the pump light before the OPO cavity and broaden the spectrum in a photonic crystal fiber (PCF) to provide spectral coverage down to 600 nm. This pump supercontinuum (*p*-SC) is superimposed with $p+s$ and $p+i$ on the respective detectors. Hence, the observed heterodyne beats correspond to $f_0^s$ and $f_0^i$, respectively. One of these frequencies is stabilized to an rf reference from a synthesizer (DDS) by feeding back to a piezo (PZT) mounted on M5; the other one may be stabilized via $f_0^s + f_0^i = f_0^p$ by feeding back to the fiber laser's pump diode power.

The resulting stability and comb tooth linewidth is determined by generating an optical beat with a 10-kHz-linewidth continuous-wave (cw) external cavity diode laser (ECDL) operating at 1.545 µm as shown in Fig. 4. The idler spectrum is tuned to a center wavelength of 3.09 µm, frequency-doubled in a GaSe crystal and combined with the ECDL light in a single mode fiber. In order to achieve better short-term stability of the OPO comb, we lock $f_{rep}^p$ (and therefore $f_{rep}^i$) in the optical domain by generating a beat note between the Yb:fiber pump laser and a 1-kHz-linewidth non-planar ring oscillator (NPRO) at 1.064 µm [see Fig. 4(a)]. With the idler tuned to 3.09 µm (and signal to 1.64 µm) the stabilization of $f_0^i$ is particularly simple because the mixing signals $p+i$ and $2s$ happen to be spectrally overlapped. Therefore, the beam focused onto PD2 contains an intrinsic heterodyne beat, and light from the PCF is no longer required. The resulting rf frequency represents $2f_0^i - f_0^s$ and is therefore an appropriate feedback for stabilizing the OPO length [see Fig. 4(b)], since the free-running $f_0^p$ shows sufficient stability (note, $f_0^s + f_0^i = f_0^p$ stable). Once the idler output is locked, a constant beat note between the doubled mid-IR comb and the ECDL is detected [see Fig. 4(c)]. The wide pedestal is consistent with the one observed in the repetition rate lock. The coherent peak in the beat signal exhibits a 3-dB-linewidth of 80 kHz, which corresponds to 40 kHz in the fundamental mid-IR comb. This linewidth is currently limited by the bandwidth of the feedback to the PZT.

In summary, we have presented a high-power OPO frequency comb that will serve as an excellent source for coherent spectroscopy techniques such as CE-DFCS in the wavelength range of 2.8 to 4.8 µm and 1.37 to 1.76 µm. The maximum idler power of 1.50 W is the highest power reported to date for a synchronously pumped OPO in the mid-infrared wavelength region. Furthermore, watt-level output power may be obtained over a significant part of the tuning range, which promises exceptional signal-to-noise ratio for spectroscopy experiments. Additionally, the covered spectral region contains strong fundamental vibrational bands of many important trace gases. Thus, detection sensitivities will be drastically enhanced compared to experiments operating in the near-infrared that have to rely on weaker overtone transitions [4,8,9]. With frequency stabilization of this mid-IR comb, we can also expect its use for future comb spectroscopy and control experiments in emerging fields such as ultracold molecules.

We thank D. C. Yost, M. J. Martin and T. R. Schibli for technical assistance and valuable discussions. F. Adler acknowledges support by the Alexander von Humboldt foundation. K. C. Cossel acknowledges support from the NSF Graduate Research Fellowship. This work is funded by AFOSR, DARPA, NIST, and Agilent.

**Figure captions**

Fig. 1  (Color online) (a) Schematic of the OPO setup. M1-M5, OPO cavity mirrors; DM, dichroic mirrors; PBS, polarizing beam splitter; PD1-2, Si photo diodes; PCF, photonic crystal fiber; DDS, rf synthesizer; PZT, piezo. (b) Snapshot of the entire spectrum emitted by the OPO from UV to IR; the arrows indicate the spectral shift for increasing crystal poling period.

Fig. 2  Idler output power (black triangles) and photon conversion efficiency (grey circles) of the OPO for different idler wavelengths. The grey dashed line represents the trend for the conversion efficiency expected from the mirror reflectivities. The inset shows the idler power versus pump power characteristics at $\lambda_i = 3.0$ μm.

Fig. 3  (Color online) Collection of typical idler spectra.

Fig. 4  Experimental setup and rf signals for stabilization of the OPO and measurement of the beat note between the doubled idler and a narrow-linewidth diode laser. NPRO, non-planar ring oscillator; YFL, Yb:fiber pump laser; ECDL, external cavity diode laser; PD2-4, photo diodes; SHG, second harmonic generation crystal (GaSe). (a) In-loop signal for stabilization of $f_{rep}^p$, 1 kHz resolution bandwidth (RBW). (b) In-loop signal for stabilization of $f_0^i$, 3 kHz RBW. (c) Out-of-loop beat note between doubled idler and ECDL, with locked OPO (black, 2 s sweep time, 3 kHz RBW) and unlocked OPO (grey, 0.5 s sweep time, 10 kHz RBW); the traces are vertically offset for clarity.

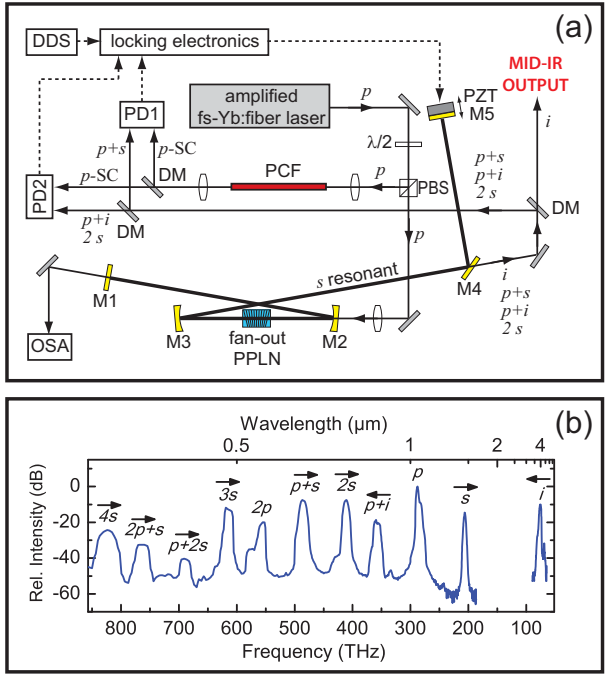

Fig. 1 (Color online)

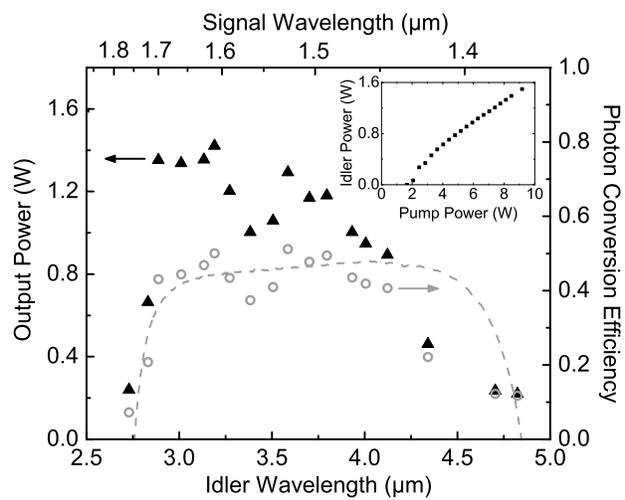

Fig. 2

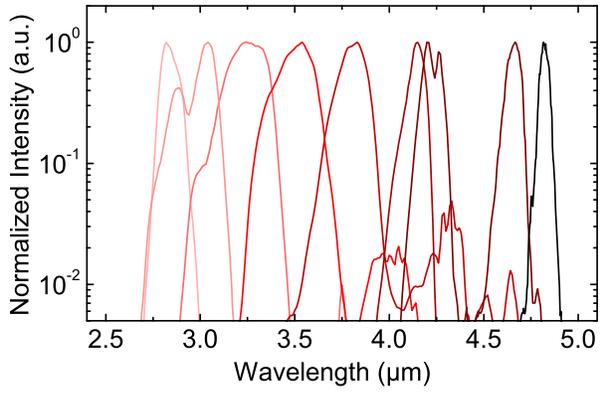

Fig. 3 (Color online)

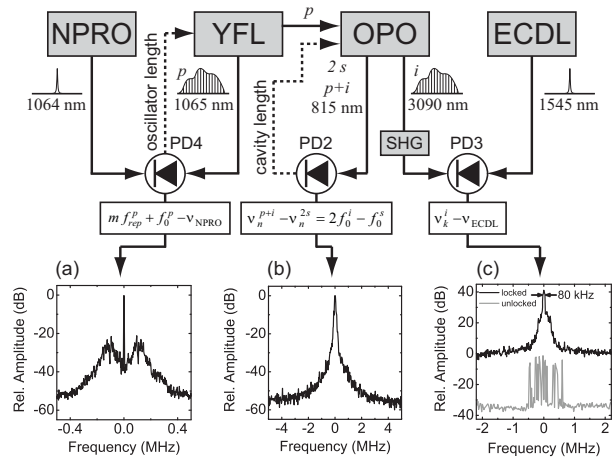

Fig. 4